\pdfoutput=1

\documentclass[conference]{IEEEtran}
\IEEEoverridecommandlockouts

\usepackage{graphicx}
\usepackage{url}

\usepackage{cite}

\ifCLASSINFOpdf
\else
\fi

\begin{document}
%
\title{Improving Block-level \\ Efficiency with \emph{scsi-mq}}

\author{\IEEEauthorblockN{Blake Caldwell}
\IEEEauthorblockA{Oak Ridge National Laboratory\\
Oak Ridge, TN\\
Email: blakec@ornl.gov}
\thanks{
This manuscript has been authored by UT-Battelle, LLC under Contract No. DE-AC05-00OR22725 with the U.S. Department of Energy. The United States Government retains and the publisher, by accepting the article for publication, acknowledges that the United States Government retains a non-exclusive, paid-up, irrevocable, world-wide license to publish or reproduce the published form of this manuscript, or allow others to do so, for United States Government purposes. The Department of Energy will provide public access to these results of federally sponsored research in accordance with the DOE Public Access Plan (http://energy.gov/downloads/doe-public-access-plan).
}
}


%


\maketitle

\begin{abstract}
Current generation solid-state storage devices are exposing a new bottlenecks
in the SCSI and block layers of the Linux kernel, where IO throughput is
limited by lock contention, inefficient
interrupt handling, and poor memory locality. To address these limitations,
the Linux kernel block layer underwent a major rewrite with the \emph{blk-mq} project to move from
a single request queue to a multi-queue model. The Linux SCSI subsystem rework to make use of this new model,
known as \emph{scsi-mq}, has been merged into the Linux kernel and work is underway for \emph{dm-multipath}
support in the upcoming Linux 4.0 kernel. These pieces were necessary to make use of the multi-queue block layer in a
Lustre parallel filesystem with high availability requirements. We undertook adding support of the 3.18 kernel to Lustre
with \emph{scsi-mq} and \emph{dm-multipath} patches to evaluate the potential of these efficiency
improvements. In this paper we evaluate the
block-level performance of \emph{scsi-mq} with backing storage hardware representative of a HPC-targerted Lustre filesystem. Our findings show 
that SCSI write request latency is reduced by as much as 13.6\%. Additionally, when profiling the CPU usage of our prototype
Lustre filesystem, we found that CPU idle time increased by a factor of 7 with Linux 3.18 and \emph{blk-mq} as compared to a standard 2.6.32 Linux kernel.
Our findings demonstrate increased efficiency of the multi-queue block layer even with disk-based
caching storage arrays used in existing parallel filesystems.
\end{abstract}


%

\section{Introduction}
The Lustre~\cite{www:lustre} parallel filesystem achieves high aggregate IO throughput for parallel applications by splitting IO requests across many backing storage devices, known as Object Storage Targets (OST's). Once file layout has been set and metadata operations have been handled by the Metadata Server (MDS), clients may send IO requests for data blocks directly to the OST. An OST handling IO requests from a large number of clients often becomes IO throughput bound due to the block storage device that backs the OST.  For assuring data integrity and reliability, these block devices are typically RAID volumes with hardware-caching capabilities that improve upon the latency qualities of the underlying magnetic hard disks. Even though the filesystem IO latency can be amortized by splitting the IO requests among many OST's, the request latency of the slowest OST will still be the determinant of overall completion time.

Hardware caches in RAID storage arrays improve both the latency and throughput within the bounds of the available cache size such that IOP rates can rival, and may even exceed the capabilities of SSD devices. The storage array used in our evaluation is capable of more than 300,000 IOP/s per RAID device when performing sequential writes. While the handling capacity of the Linux block subsystem is limited to somewhere between 400,000~\cite{assche15} and 800,000 IOP/s~\cite{bjorling13}, a penalty is paid as IOP/s near the upper bound of that range in terms of increased lock contention, cache invalidations, and interrupt servicing. However, when a multi-queue block layer is employed, as in~\cite{bjorling13}, Linux itself can scale to over 10~million~IOP/s.

Previous performance evaluations of \emph{blk-mq} have been done using the \emph{null-blk} device driver~\cite{assche15, bjorling13}, which simply acknowledges the IO operation, without writing to the data to any storage device~\cite{bjorling13}. We build upon these evaluations, but take a different angle of evaluating caching storage arrays such as those providing OST's for production Lustre filesystems. To demonstrate the applicability of block-level throughput and latency improvements, we have built a Lustre filesystem on the 3.18 Linux kernel patched with \emph{scsi-mq} support.

The remainder of this paper is structured as follows. In Section~\ref{sec:back}, we describe the Linux block subsystem and block-layer enhancements. Section~\ref{sec:eval} describes our prototype filesystem and and experimental setup. In Section~\ref{sec:results}, we present the results from our block-level throughput and latency tests and kernel code profiling with \emph{scsi-mq}. Then Section~\ref{sec:discuss} discusses our findings on block-layer improvements in the context of a Lustre filesystem. We present related works in Section~\ref{sec:related} and we conclude with Section~\ref{sec:concl} which summarizes our findings and areas for future work.

\section{Background}
\label{sec:back}
\subsection{Legacy block layer}

Even prior to the \emph{blk-mq} update, the Linux block layer was adequate for magnetic hard disks with slow seek latencies, where IOP rates weren't constrained by lock contention or interrupt handling. Rather than optimizing the block layer for minimum latency, it was often even advantageous to leave the queue in a ``plugged'' state where requests sat in the queue waiting to be coalesced with other requests into a single sequential IO. With rotational drives, the delay introduced by request merging and reordering is far outweighed by the reduction in seek latencies by making IO sequential.

However, with SSDs capable of servicing millions of IO operations per second~\cite{zheng13}, the tradeoffs made by the existing Linux block layer needed to be re-evaluated. 
The designers of \emph{blk-mq} detailed three major
scalability concerns in the legacy Linux block layer in~\cite{bjorling13}. Firstly, insertions, removals and IO scheduling
operations involving the request queue require obtaining a global lock, where cores competing for
the same lock will block waiting for the operation to complete.
Secondly, beyond just one hardware interrupt per IO completion, the exact core handling the IO request may still need to be signaled by an inter-processor interrupt (IPI).
Thirdly, cache coherency must be maintained. Multiple cores must contend for the per block device
request queue lock. Each transfer of ownership of the lock between cores will cause cache line
invalidations and cache coherency traffic across the memory bus. If the lock remains in memory on
within a different NUMA locality region, then acquiring the lock will involve a remote memory
access.

Because of these effects, Bj{\o}rling et al. find that the Linux block layer is limited to 800,000 IOP/s, and cannot benefit from the use of additional cores~\cite{bjorling13}.


\subsection{Multi-queue enhancements}

To address the three concerns outlined in~\cite{bjorling13} and restated above, the block layer was restructured in the Linux 3.13 kernel with two levels of queues.
The \emph{blk-mq} design employes per-core software submission queues, where an IO request is enqueued on the same core as the application submitting the IO.
The queue obeys simple FIFO semantics, removing the need for locking of the queue for each insertion or removal. Additionally, accessing the queue structure 
is done entirely within the memory locality region of the core submitting the request. A second level of queues has the responsibility of dispatching IO's
to the underlying storage device driver. These queues are usually equal in number to the degree of parallelism supported by the device. For example, SSD drives can exploit the parallelism of independent NAND flash chips. Some PCIe SSD devices have up to 32 channels per controller~\cite{Cornwell:2012} while SATA-attached SSD's have up to 10 channels~\cite{www:x25spec}. For legacy block device drivers such as those for rotational hard disks, 
the number of hardware dispatch queues defaults to one.

While the legacy block layer takes advantage of time an IO request remains in the submission queue to sequentially reorder requests, the \emph{blk-mq} design simply interleaves IO requests from the per-core software submission queues onto the hardware dispatch queues. With solid state devices, capable of higher random IO performance, the benefit from sequential ordering is lessened and the FIFO queues avoid locking overheads that become more
pronounced at high IOP/s rates. This trade-off may be troublesome for rotational media, but as noted in~\cite{bjorling13}, there is room for IO scheduling to be done between the two queue levels. To date, this scheduling has not been implemented.

Another enhancement added by \emph{blk-mq}, is support for tag-based completions.  The block layer associates a tag for
each IO, where the tag corresponds to the position in the hardware dispatch queue. The tag is then preserved in the device driver code and can be used again by the block layer
to identify the IO upon completion. This increases the speed at which the IO completion can be identified and then passed on for handling by the core
that was responsible for its submission.

\begin{figure}[!t]
\centering
\includegraphics[width=6.5cm]{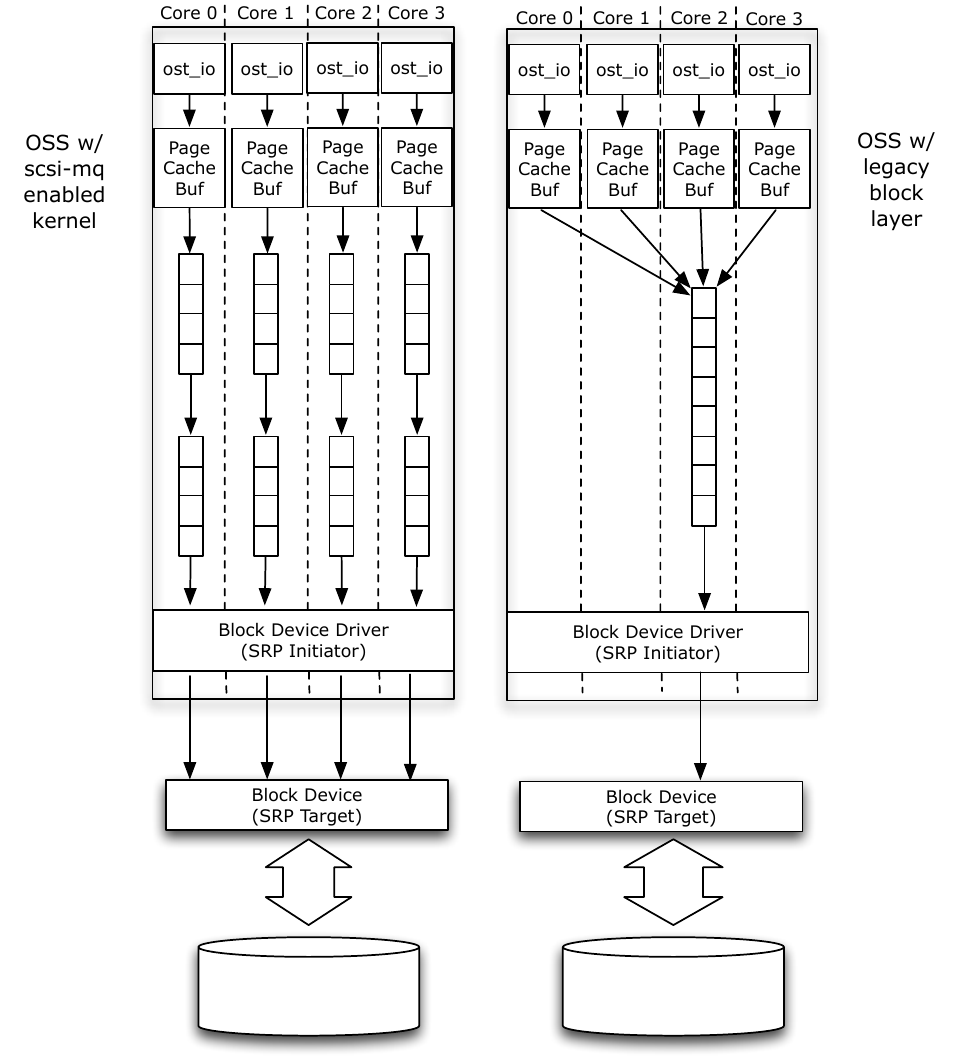}
\caption{A multi-queue block layer (left) vs. single queue block layer (right)}
\label{fig_scsi-mq}
\end{figure}

The multi-queue block layer design is shown on the left in Figure~\ref{fig_scsi-mq}, where \emph{ost\_io} threads on each core submit IO's through the Linux kernel page cache to a dedicated submission queue. These submission queues map to an equal number of dispatch queues supported by the SRP initiator device driver, with multiple RDMA channels between the OSS and the SRP target implemented in the storage array. This design contrasts to the right half of Figure~\ref{fig_scsi-mq} with a single queue per block device, where \emph{ost\_io} threads on different cores are competing for the same lock.
 
The performance reported with \emph{scsi-mq} has ranged from 1 million IOP/s~\cite{hellwig14} using an array of 16 SAS SSD's to 12 million using the \emph{null-blk} device driver which just acknowledges the request and discards the data~\cite{bjorling13}.

\subsection{SCSI subsystem work}
To make use of the multi-queue block layer with SCSI devices, the SCSI IO submission path had to be updated. The \emph{scsi-mq} work was
functional in the 3.17 kernel and lower level device drivers are being updated to support multiple hardware dispatch queues.
Of interest to this study, is the SRP initiator driver \emph{ib-srp} which was updated in the Linux 3.19 kernel (and in our patched 3.18 kernel) 
to support \emph{scsi-mq} and multiple RDMA channels~\cite{assche15}.

Other work is underway on the \emph{dm-multipath} driver to support \emph{scsi-mq}. This feature is highly desirable for Lustre deployments and the
\emph{dm-mulitpath} maintainers are targeting inclusion of an updated multi-queue compatible driver in the upcoming Linux 4.0 kernel release

\subsection{Lustre IO patterns}
\label{sec:lustre-io}
A common measure of the performance of parallel filesystems in HPC is the
aggregate IO handling capacity, and to reach near 1~TB/s~\cite{sc14-spider}, parallelism is used at multiple levels.
Lustre stripes objects across multiple block storage devices \cite{www:lustre}
and each block device is backed by a RAID volume that stripes its data across multiple
hard disks. Large IO requests can benefit from this parallelism, roughly by a
factor equal to the stripe count. However each
operation incurs a certain amount of overhead associated with the metadata
tracking, ordering, and consistency checking, where small IO requests are
disproportionately affected by individual request latency. Lustre metadata operations,
such as a \emph{stat} typically involve IO requests 4 KB in sizem
and bulk IO operations to the OST's are ideally done
in 1 MB segments.

It is not immediately apparent how the large 1 MB IO requests to Lustre OST's 
from would benefit from either increased IOP/s capacity or decreased request
latency. In fact prior studies of multi-queue block-layer performance~\cite{bjorling13, www:scsimq:v2}, have used
512-byte sized IO's to the reach high the IOP/s rates where \emph{blk-mq} shines. With 1 MB IO's, an IOP/s rate of just 1024
is required to reach 1GB/s of throughput. The streaming throughput capacity of the RAID volume will become the bottleneck before the 
legacy Linux block layer in this case. A reduction in request latency is unlikely to be noticeable
with 1 MB object storage requests because a data transfer rate of 1 GB/s means that latency will be roughly 1 ms, compared
to the few $\mu$s difference in IO handling latency.

However, for 4KB IO operations, IOP/s rates begin to approach what the Linux block
layer can handle. The vendor of the storage
hardware used in this study claim that the array we used is capable of 840,000~IOP/s~\cite{ddn10k} out of the cache.
While we were only able to attain 500,000~IOP/s at this block size,
such a rate is still approaching the limitations of the Linux block layer~\cite{bjorling13}. Additionally, request latencies of
these small IO's becomes important for the interactivity of the filesystem to the end-user.


\section{Evaluation}
\label{sec:eval}
Our prototype filesystem for evaluating the performance impact of \emph{scsi-mq}
was built using hardware similar to known production filesystems~\cite{sc14-spider}. The testbed was built around
a DataDirect Networks SFA 10K-X storage appliance housing 600 2TB SATA disks. 
The disks are organized in 60 RAID-6 pools of 10 disks each. The pools
are each presented as a block device via SCSI RDMA Protocol (SRP) over InfiniBand to
four Lustre object storage servers (OSS). Typically four OSS  are needed to
meet the bandwidth capacity of the storage array, but only one OSS was used
for this evaluation. We tested cases of multiple block devices, or Logical Unit Numbers (LUNs),
attached to a single host, but not the case of using multiple OSS.

Each OSS has two
Intel E5-2650 Ivy Bridge processors, 64GB RAM, and 2 QDR InfiniBand links for the SRP
transport to the storage array. Using \emph{dm-multipath}, and the dual InfiniBand links between
the OSS and the storage array, highly available block devices could be used, which would
survive the failure of one of the two storage controllers in the storage array. Since the
\emph{dm-multipath} drivers add an abstraction layer above the block-device, which could
negatively affect the performance, we set out to quantify the performance penalty, and then
use the raw block device to for the remainder of the block-level tests. The patches to enable
\emph{dm-multipath} to work with \emph{blk-mq} were added to the 3.18 kernel used in this
evaluation. These patches are proposed for inclusion in the Linux 4.0 kernel.

Since we are studying the performance capabilities of the Linux block device layer, we
turned on write-back caching on the storage array. We acknowledge that the
sustained IOP rates of each rotational drive-backed RAID volume would be significantly
lower and thus become the bottleneck in tests that exhausted the array's cache capacity of 8 GB
mirrored between the controllers. The storage array has an additional cache setting called ReACT, which disables the write-back caching
for aligned full-stripe writes~\cite{ddn10k}. We left this feature enabled, as is recommended by the 
vendor for use with production filesystems, but it was not impactful to the tests run in this evaluation
using 4 KB IO's since the array's configured stripe width is 1MB.

In keeping with the goal of evaluating \emph{scsi-mq} performance in an
environment representative of current Lustre
filesystem deployments, different software stacks were deployed on the OSS
in using a diskless provisioning technique. The complete configuration 
and OS install are captured in system image that differs only where necessary between 
the software stacks evaluated. Three images were constructed: one with Red Hat
Linux 6.6, and kernel version 2-6.32-431, and the other two with Red Hat Linux
7.0 with kernel 3.18. The only difference between the two Red Hat 7.0 images
is the setting of the kernel configuration parameter
\emph{CONFIG\_SCSI\_MQ\_DEFAULT}. With this setting enabled, in addition to per-core software
queues, the SRP lower-level SCSI device driver made use of multiple hardware dispatch queues and
equal to the number of SRP channels configured. Disabling this parameter meant that a single hardware
dispatch queue would be used, but with multiple per-core software queues. These scenarios
with the 3.18 kernel are referred to as \emph{mq+mc} and \emph{mq} respectively in Section~\ref{sec:results}.

The SRP transport protocol protocol involves drivers on both the OSS (SRP initiator) and
storage array (SRP target).  The SRP target driver has supported multi-channel operation previously,
but recent patches to the SRP initiator driver \emph{ib\_srp} from Bart Van
Assche added options to configure the number of channels and set the
completion request affinity. In this study, we tried two different channel configurations. First, with 16 channels, one per core,
and then two channels, with one per processor socket. We did not observe a significant
difference in performance when using 16 compared to two channels, so for the results
presented below, we used 16 channels for comparison to other studies that used one channel
per core~\cite{www:scsimq:v2}.

At the time of completion of this study, support for the 3.18 kernel was not yet
part of the Lustre code. Since it was a primary objective of this study to determine
what performance benefits \emph{scsi-mq} could bring to Lustre, we
made the necessary changes to the Lustre code to be compatible with the
patched 3.18 kernel which included \emph{scsi-mq} support. One of the changes
required was rebasing patches from the upstream Lustre client driver to the master
branch of the Intel Lustre tree. Many production Lustre filesystems use a special filesystem
format called \emph{ldiskfs} for the object and metadata target devices. Unlike
other backing filesystems such as \emph{zfs}, which are less-commonly used with Lustre,
especially for metadata targets, \emph{ldiskfs} is a series of patches on
top of the \emph{ext4} filesystem. These patches were rebased for the \emph{ext4}
code in the Linux 3.18 kernel for this study. We have made all of our patches to
Lustre for the 3.18 kernel publicly available on GitHub~\cite{www:github:lustre318}.

As with other works involving storage performance on NUMA architectures
\cite{zheng13, shelton13}, we had to pay careful attention to
processor and interrupt affinity in setting up the experiments. The InfiniBand card connecting
the OSS to the storage array has affinity to NUMA 
node~1, so we directed all interrupts from the Infiniband device to the cores on the single
processor socket which makes up NUMA node~1. Also \emph{numactl} was
used to bind the workload generation processes to the cores on node~1.

The workload in each test was generated with \emph{fio} \cite{www:fio} using the libaio IO engine.
Parameters for each of the tests were a
runtime of 60 seconds, iodepth of 127, and IO's batched in units of 62. The average of tree trials for each test
is presented in the figures in Section~\ref{sec:results}.
For the SRP initiator parameters, we set max\_cmd\_per\_lun to 62, and queue\_size to 127.  We ran all block-level
tests with a workload of 4 KB sequential writes, representative of a specific use case in a Lustre filesystem,
metadata IO. Writes were chosen instead of reads for an initial in-depth study, to avoid the need to account for
the effects of a read-ahead prefetch behavior. An evaluation of read performance is a subject we aim to include in
future work. We only studied sequential performance with the rotational disk-backed storage array, 
because performance with random IO's involving large seek times drops to a level such that the 
bottleneck was never the Linux block layer.
While this would be a useful measure or performance on a filesystem
backed by flash storage media, capable of high random IOP/s rates, it is not applicable to our study
with rotational media.

\section{Results}
\label{sec:results}

\begin{figure}[!t]
\centering
\includegraphics[width=6.5cm]{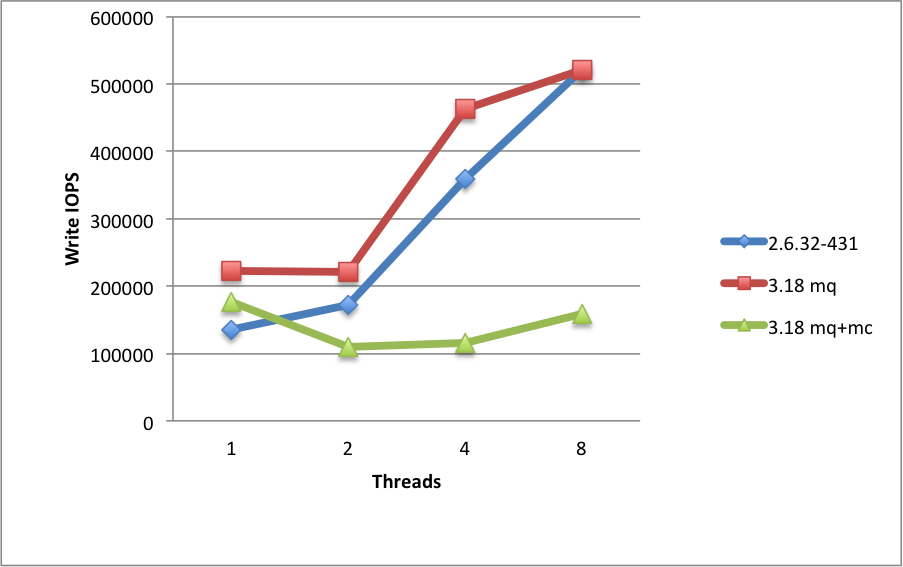}
\caption{4 KB Sequential writes on a single LUN}
\label{fig_threads}
\end{figure}

As a first point of comparison, Figure~\ref{fig_threads} shows the throughput effects by increasing the number of
\emph{fio} threads submitting IO's to a single block device. The Linux 3.18 stack with multiple per-core submission
queues has an edge over the 2.6.32-431 kernel with a single submission queue from 1 to 4 threads per LUN.
However, the values converge at 8 threads per LUN. We expect that lock contention is greatest in the 8 thread per
LUN case, which should have benefited the 3.18 \emph{mq} stack, but it does not appear to be the bottleneck since
the legacy block layer is able to equal IOP/s at just over 500,000. As is the case throughout our evaluations, the
3.18 kernel with multiple SRP channels fails to show any benefit from increased parallelism. We did not have
the opportunity as part of this study to determine what is preventing the \emph{mq+mc} stack from scaling, but
this is an area we intend to study in future work.

\begin{figure}[!t]
\centering
\includegraphics[width=6.5cm]{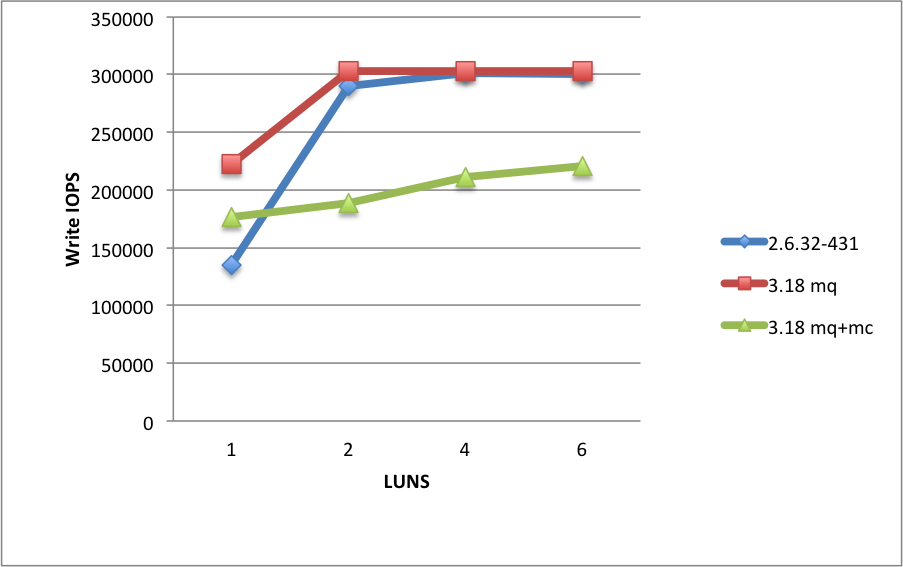}
\caption{4 KB sequential writes with a single thread per LUN}
\label{fig_luns}
\end{figure}

An unexpected finding in our tests was that the performance of multiple threads per LUN exceeded that of writing to multiple LUNs. In Figure~\ref{fig_luns},
the throughput of both the legacy block layer and 3.18 \emph{mq} tops out near 300,000 IOP/s. This hints that a bottleneck
exists somewhere outside the factors addressed by \emph{blk-mq}. The performance of multiple threads writing to a single block
device in Figure \ref{fig_threads} exceeds the 300,000 IOP/s mark.  We would need to perform additional analysis to determine
whether that is a result of cache optimizations, such as request reordering either the block level or within
the storage array hardware.

As discussed in Section \ref{sec:lustre-io}, reduced request latencies for 4 KB IO's could have a positive impact on filesystem
interactivity. We list the average latency of 100 timings of 4 KB requests with the pre-\emph{scsi-mq} kernel and a Linux 4.0 release candidate
with the \emph{dm-multipath} patches for \emph{scsi-mq} in Table~\ref{tab:latency}. A significant 13.6\% improvement in 4 KB write latency is seen
with the \emph{scsi-mq} kernel.  The request latency with \emph{dm-multipath} remained the same between the two software stacks, however.

\begin{table}[!h]
\caption{4KB Write Latency ($\mu$s)}
\centering
\begin{tabular}{l c c}
\hline\hline
{\bf Kernel} & {\bf dm-multipath} & {\bf raw block} \\
\hline
   2.6.32-431 &   130.50  &   119.92 \\
   4.0rc1 &   130.56  &   103.58 \\
\hline\hline
\end{tabular}
\label{tab:latency}
\end{table}

\begin{figure}[!t]
\centering
\includegraphics[width=7.5cm]{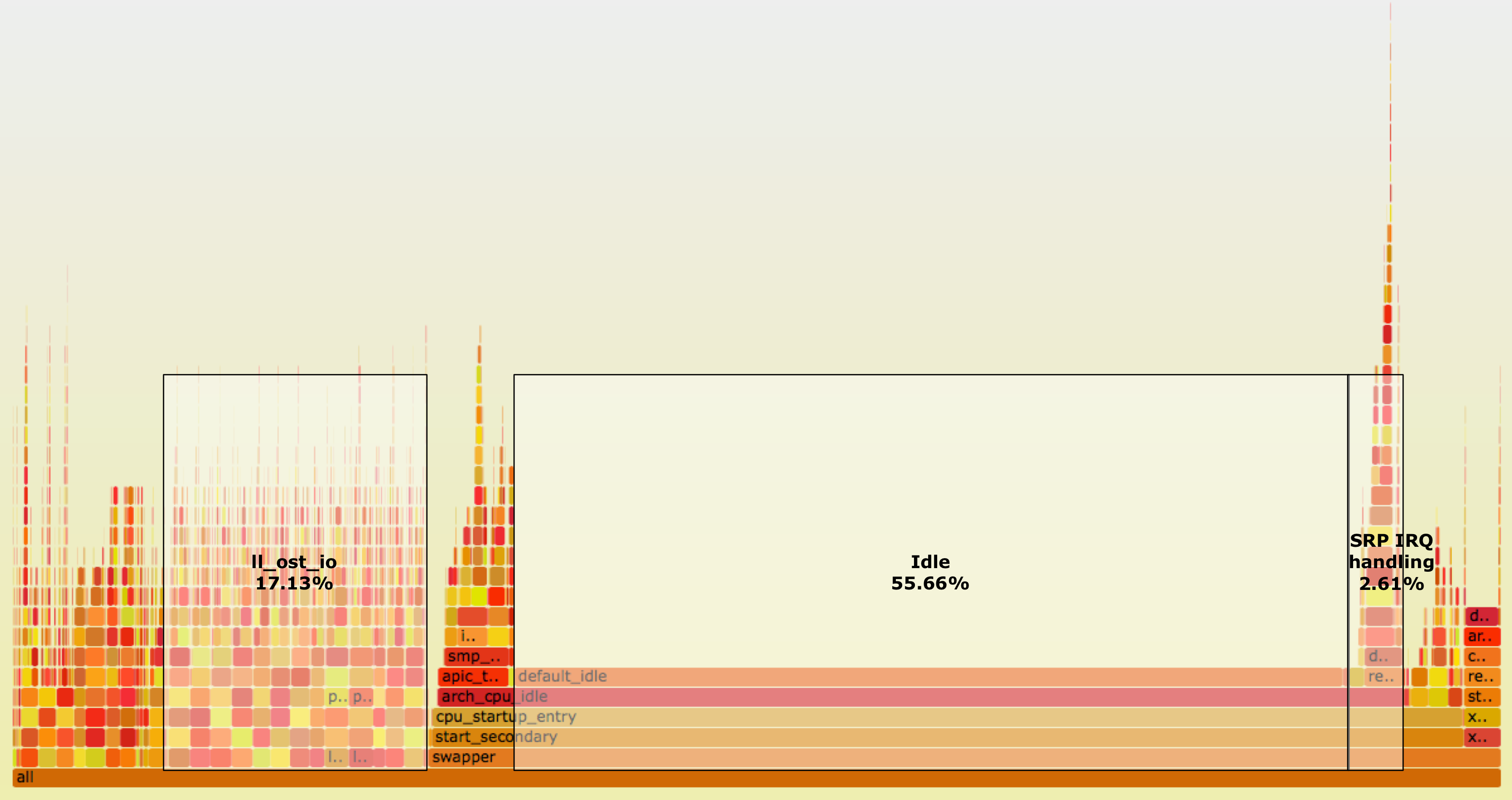}
\caption{Sampled kernel stack traces of Lustre OST writes with scsi-mq and a 3.18 kernel}
\label{perf-mq}
\end{figure}

\begin{figure}[!t]
\centering
\includegraphics[width=7.5cm]{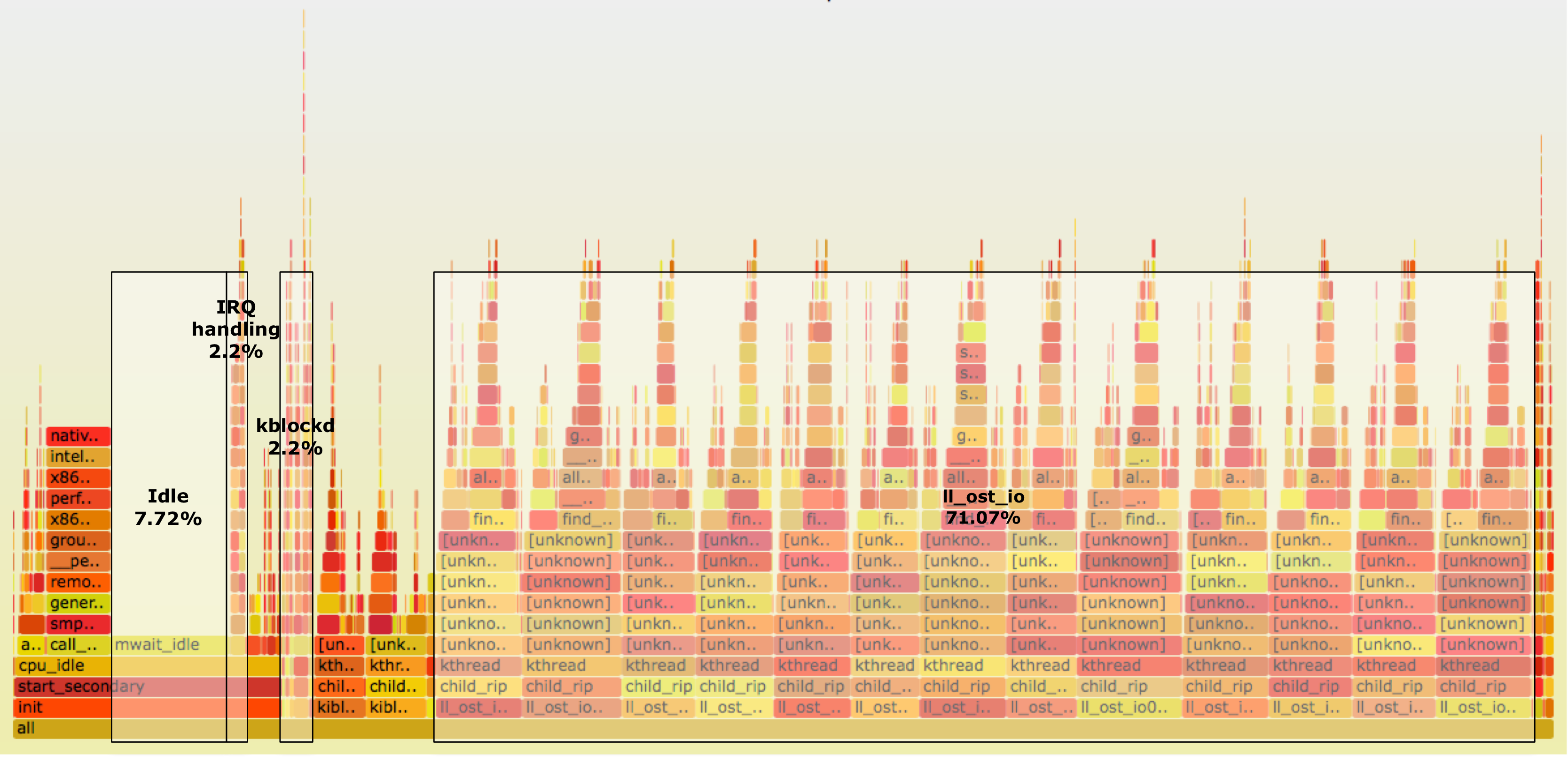}
\caption{Sampled kernel stack traces of Lustre OST writes with a 2.6 kernel}
\label{perf-2.6}
\end{figure}

In addition to latency and throughput performance metrics, we looked for signs of increased efficiency at the filesystem-level. Using the
\emph{perf} tool, kernel stack traces were collected at 100 Hz, and compiled into FlameGraphs \cite{flamegraphs} shown in Figure \ref{perf-mq} 
with \emph{scsi-mq} and \ref{perf-2.6} with the legacy block layer.  A striking difference is seen the relative amount of cycles spent idle vs.
in Lustre \emph{ll\_ost\_io} threads. With the 2.6 Linux kernel, 7.72\% of cycles are spent idle, while 55.06\%
of sampled cycles are idle cycles with the 3.18 kernel. The increase in idle cycles can be attributed in large part to decreased time spent in \emph{ll\_ost\_io}
threads while handling IO requests and their completion.

\section{Discussion}
\label{sec:discuss}

The workloads evaluated in this study are focused on the use case of Lustre metadata operations, which have an IO size of 4 KB. These small
IO requests have higher sensitivity to changes in latency, and thus the benefit from a 13.6\% reduction in block-level request latency
could benefit a real filesystem-level workload. One such benefitting workload would be a recursive directory listing that gathers the total size of a files contained
underneath a directory. There is a data dependency between reading the contents of a directory entry, and subsequent reads of each inode
part of the directory entry. These data dependencies force a synchronous order of metadata operations where the completion latency
of each one directly impacts the overall delay a user would experience. Filesystem scans
that walk the whole directory structure also encounter a data dependency in metadata operations, and could see a benefit from reduced 4 KB
request latency.

Metadata throughput of a single block device in Lustre has not been not been measured at a rate near the IOP
capacity of the single request queue block layer. An OpenSFS-funded project \cite{www:mdssmp} measured
up to 200,000 IOP/s during concurrent \emph{stat} operations by 16 clients, but improvements in subsequent Lustre
versions could have pushed the IOP/s rate close to or above the handling capacity of a single block device without
\emph{scsi-mq}. An example of an event that would trigger high IOP/s rates to the block device would be cache
write-back events on the MDT.  A large stream of cache write-backs would occur at less interesting times such as 
when unmounting a filesystem, but perhaps there a more comprehensive study of metadata target block IO might
reveal other times. Another time the MDT might become IOP-limited is during filesystem recovery when the journal is
being played back and committed to the filesystem.

A different scenario where the multi-queue block layer design could be beneficial is when multiple IO's are being issued to
the same block device, but one is more latency sensitive that the rest. In fact,
a benefit mentioned in \cite{bjorling13} of \emph{blk-mq} was that 2-level queuing allows for IO scheduling, and 
potentially a quality of service mechanism to intercede as IO's from the per-cpu software queues are inserted into
the hardware dispatch queues. Reserving queue resources for smaller or more-latency sensitive IO operations
could lead to predictable latency variation during times of high load. Furthermore, the ability of
\emph{scsi-mq} to preserve IO request tags used in the block layer into the SCSI device driver level, and 
on completion of the IO adds an accounting mechanism for the timing of individual IO requests through completion.

\section{Related Work}
\label{sec:related}

The \emph{blk-mq} design was detailed by Bj{\o}rling et al. in~\cite{bjorling13}, and then merged in Linux 3.13 kernel.
In order to add support of this new IO path to SCSI subsystem, the \emph{scsi-mq} work was merged in the 3.17 kernel~\cite{hellwig14}.
Each of the works ~\cite{bjorling13, hellwig14}, contributed performance evaluations, with the former utilizing the \emph{null-blk}
device driver for measuring pure software overheads and the latter using an SSD-back RAID array.  The results in~\cite{hellwig14}
have applicability to some real Lustre filesystem deployments, but many of the highest performing Lustre filesystems make use of
a RAID array and the SRP transport which Bart Van Assche added multi-queue support to in Linux 3.19~\cite{assche15}.  Van Assche
also measured the performance of \emph{scsi-mq} before the 3.17 merge and published results using the \emph{null-blk} driver in \cite{www:scsimq:v1, www:scsimq:v2}, and
detailed SRP multi-queue status in~\cite{www:scsimq:ofa, assche15}.

While previous evaluations examined the software efficiency of \emph{blk-mq}, our work examines the impact on
performance with a caching storage array and representative parallel filesystem IO patterns. This is, to our knowledge,
the first evaluation of parallel filesystem making use of the multi-queue Linux block layer.

\section{Conclusion}
\label{sec:concl}

In this paper we have detailed our experimental evaluation of a prototype Lustre filesystem using the redesigned
multi-queue block layer in a patched Linux 3.18 kernel. We have made our Lustre code modifications for the Linux 3.18
kernel with \emph{scsi-mq} publicly available on GitHub~\cite{www:github:lustre318}. Our block-level tests revealed
a 13.6\% improvement in write request latency
over the standard block layer in the Linux 2.6.32 kernel and that idle CPU cycles increased by a factor of 7.
Within the context of a Lustre filesystem, we explored benefits that these improvements could bring to overall
filesystem performance. While these results are promising, this study has also highlighted areas in need of further
investigation.

As a subject of study in future work, we intend to understand the reasons behind the poor scaling
performance of multi-channel support in the SRP driver. We found performance improvements
with the use of per-core software submission queues, but utilizing multiple hardware queues actually degraded
performance. This is contrary to what we expected given the performance benefits of a
multi-channel SRP driver in \cite{www:scsimq:v2}.  Additional
channels should expose additional parallelism in the storage array hardware and avoid contention for
shared queues.  We focused on write performance in this paper, but an in-depth investigation of read
performance is also necessary to understand the full impact of the redesigned block layer.

We have narrowed our target for filesystem improvements from \emph{scsi-mq} to metadata operations
where IOP rates near the capacity of the existing Linux block layer, and requests are sensitive to latency improvements.
The caching storage arrays with rotational media used in this study were capable of IOP rates near the cited limits of the single request
queue block layer, but we were unable to verify the observed IO throughput was limited by inefficiencies in the
software stack. Performing evaluations on solid-state media capable of higher IOP/s rates and high random
throughput could highlight these limitations in future work.

\ifCLASSOPTIONcompsoc
  \section*{Acknowledgments}
\else
  \section*{Acknowledgment}
\fi

This research used resources of the Oak Ridge Leadership Computing Facility at the
Oak Ridge National Laboratory, which is supported by the Office of Science of the 
U.S. Department of Energy under Contract No. DE-AC05-00OR22725.


\bibliographystyle{IEEEtran}
\bibliography{IEEEabrv,bib/paper}
%



\end{document}